\documentclass[12pt]{iopart}
\usepackage{iopams}  
\usepackage{graphicx}
\usepackage{color}
\usepackage[normalem]{ulem}

\begin{document}
\title []{Does the CMB prefer a leptonic Universe?}
\author{Dominik J Schwarz$^1$ and Maik Stuke$^{1,2}$}
\address{$^1$ Fakult\"at f\"ur Physik, Universit\"at Bielefeld, Postfach 100131, 33501 Bielefeld, Germany.\\
$^2$ Gesellschaft  f\"ur Anlagen und Reaktorsicherheit, Forschungszentrum Boltzmannstr. 14, 85748 Garching b. M\"unchen
, Germany.}
\eads{\mailto{dschwarz} and \mailto{mstuke}  at \mailto{physik.uni-bielefeld.de}}

\begin{abstract}
Recent observations of the cosmic microwave background (CMB) at smallest angular scales and 
updated abundances of primordial elements, indicate an increase of the energy density and the 
helium-4 abundance with respect to standard big bang nucleosynthesis with three 
neutrino flavour. This calls for a reanalysis of the observational bounds on neutrino chemical potentials, 
which encode the number asymmetry between cosmic neutrinos and anti-neutrinos and thus measures 
the lepton asymmetry of the Universe. We compare recent data with a big bang nucleosynthesis 
code, assuming neutrino flavour equilibration via neutrino oscillations before the onset of big bang
nucleosynthesis. We find a preference for negative neutrino chemical potentials, 
which would imply an excess of anti-neutrinos and thus a negative lepton number of the Universe. This 
lepton asymmetry could exceed the baryon asymmetry by orders of magnitude.  
\end{abstract} 
\vspace{2pc}

\maketitle
%%%%%%%%%%%%%%%%%%%%%%%%%%%%%%%%%%%%%%%%%%%%%%%%%%%%%%%%%%%%%%%%%%%%%%%
%
%
\section{Introduction}

Two cornerstones of modern cosmology are the measurements of the abundance of primordial light 
elements and the observation of the cosmic microwave background (CMB) radiation. Both are 
described very well by the hot big bang model. The abundance of light elements is inferred from the 
observation of  carefully selected astrophysical objects, for example extragalactic HII regions to determine 
the primordial helium abundance. The cosmic microwave background is measured for example via 
satellites like the Wilkinson Microwave Anisotropy Probe (WMAP) at large scales and with telescopes like 
the Atacama Cosmology Telescope (ACT) and the South Pole Telescope (SPT) at small angular scales. 
Both types of observation provide comparable results for the baryon density of the Universe from very 
different epochs.

Recent observations with ACT and SPT allow us for the first time to also estimate the cosmic helium 
abundance from the CMB \cite{Dunkley:2010ge,Sievers:2013wk,Keisler:2011aw,Hou:2012xq}. 
The measurement of light element abundances at late times (today) as compared 
to the epoch of photon decoupling allows us to refine the tests of standard cosmology. The recent CMB data 
from small angular scales also allow us to compare an estimate of the number of relativistic degrees of 
freedom  at the time of photon decoupling with an estimate of that number at the time of big bang 
nucleosythesis (BBN). 

For the determination of the cosmic abundance of helium-4, the CMB analysis might have an advantage 
over the measurement of extragalactic HII regions, since it is just one global dataset and there was no 
chemical evolution at the time of photon decoupling. However, the limits on the primordial abundance of 
helium are much tighter from stellar observations, but in turn the baryon density of the Universe is much 
better constrained by CMB experiments. The best dataset to describe the primordial abundance of light 
elements is a combination of both. 

In this work, we re-investigate the possibility of non-standard big bang nucleosynthesis, based on 
SPT results \cite{Keisler:2011aw, Hou:2012xq}, the final WMAP analysis 
\cite{Hinshaw:2012fq}, and the recent reinterpretation of the helium-4 and deuterium abundance 
\cite{Izotov:2010ca,Aver:2011bw,Olive:2012xf,Pettini:2012ph}. We use stellar observations and CMB data 
to constrain the influence of a possible neutrino or lepton asymmetry. To do so, we compare and 
combine different results for the abundance of primordial light elements with theoretical expectations 
including neutrino chemical potentials. We assume that neutrinos are Dirac fermions and that 
they are relativistic before and at the epoch of photon decoupling, i.e. $m_{\nu_i} < 0.1$ eV, $i = 1,2,3$.

\section{Large lepton asymmetry} 

The only free cosmological parameter of standard big bang nucleosynthesis (SBBN) is the baryon to 
photon density ratio. It is commonly defined as the difference of the number density of baryons $n_b$ and 
anti-baryons $n_{\bar b}$ normalized to the number density of the photons 
$n_{\gamma}$:  $\eta_b=(n_b-n_{\bar b})/n_{\gamma}$. 
The observed $\eta_b={\cal O}(10^{-10})$ shows a tiny excess of baryons, thus a baryon asymmetry. 

SBBN ignores a  possibly large lepton asymmetry, hidden in the three active neutrino flavours. The 
common model assumption is that sphaleron processes equilibrated the total lepton and baryon 
asymmetry in the very early universe and neutrino oscillations result in the equilibration of any lepton 
flavour asymmetry. Together both assumptions result in a tiny, unobservable lepton asymmetry. 

However, the existence of sphaleron processes has not been established by experiment so far and 
numerous theoretical models predict a significantly larger lepton (flavour) asymmetry, 
c.f.~\cite{Linde:1976kh,MarchRussell:1999ig,McDonald:1999in}. This provides motivation enough 
to consider a scenario with large lepton asymmetry. In previous works we have investigated the effects 
of large lepton asymmetry on the cosmic quark-hadron transition and the freeze-out abundance of 
weakly interacting massive particles (WIMPs) \cite{Schwarz:2009ii,Stuke:2011wz}. 
The purpose of this work is to inspect the consequences for CMB and BBN predictions.  

BBN predictions would then be modified by including neutrino flavour chemical potentials $\mu_{\nu_f}$, 
with $f=e$, $\mu$, $\tau$. For fixed temperature, the introduction of a chemical potential increases the energy 
density of neutrinos. Introducing these additional energies leads to a faster expansion of the early universe 
(see e.g.~\cite{Steigman:2007xt}).
We denote the Hubble expansion rate with non-vanishing neutrino chemical potentials by $H^{\prime}$ and 
$H$ is the Hubble rate without lepton asymmetry. The difference is commonly expressed via the 
expansion rate factor $S=H^{\prime}/H=(\rho^{\prime}/\rho)^{1/2}$, with the corresponding energy 
densities $\rho$ and $\rho^{\prime}$. The difference in the energy densities is the observed extra radiation 
energy density, commonly expressed as additional neutrino flavour in the effective number of neutrinos 
\begin{equation}
\Delta N_{\rm eff} = 
(N_{\nu} - 3) + \sum_f{\frac{30}{7}\left(\frac{\xi_f}{\pi}\right)^2+\frac{15}{7}\left(\frac{\xi_f}{\pi}\right)^4}, 
\label{DNeff}
\end{equation}
with $N_{\nu}=3$ for the three neutrino flavour $f=e$, $\mu$, $\tau$, and corresponding neutrino chemical 
potentials  $\xi_f=\mu_{\nu_f}/T_{\nu}$ at neutrino temperature $T_{\nu}$. 
Note that the standard model predicts $N_{\rm eff}=3.042$, a small excess above $3$ due to corrections 
from electron-positron annihilation (not included in (\ref{DNeff}), 
but taken into account in our numerical calculations below).
The expansion rate factor becomes $S=\sqrt{1+(7\Delta N_{\rm{eff}})/43}$. Assuming neutrino flavour 
equilibration through neutrino oscillations before the start of BBN ensures $\mu_e=\mu_{\mu}=\mu_{\tau}$ at 
$T=T_{\rm BBN}$ \cite{Mangano:2011ip}. Additional effective degrees of freedom could also 
be caused by other reasons, for example a variation of the gravitational constant 
might have a similiar effect \cite{Starkman:1992uq}. We will concentrate here only 
on neutrino asymmetry induced chemical potentials. 
Assuming relativistic neutrinos and a lepton asymmetry much bigger than the baryon asymmetry $|l|\gg b$, 
but still $|l| \ll 1$, one can link the neutrino chemical potentials to the lepton asymmetry $l$ 
\cite{Schwarz:2009ii},
\begin{equation}
\xi_f=\frac{\mu_{\nu_f}}{T_\nu} = \frac{1}{2} l \frac{s}{T^3},
\end{equation}
where $s$ denotes the entropy density.

A large lepton asymmetry leads also to a second effect during BBN, due to interactions of 
electron neutrinos with ordinary matter. While all three neutrino flavour chemical potentials affect the 
Hubble rate independently of their sign, the electron neutrino chemical potential influences the 
beta-equilibrium $e+p\leftrightarrow n+\nu_e$ directly. It shifts the proton-to-neutron ratio, depending 
on the sign of $\mu_{\nu_e}$, and so modifies the primordial abundances of light elements.
 
The two effects can be played against each other \cite{Olive:1991ru} in a way that one compensates the 
other one. For further discussions of the impact neutrinos have on BBN and CMB c.f. 
\cite{Khlopov:1981nq, Zeldovich:1981wf, Kohri:1996ke, Hansen:2001hi, Dolgov:2002wy, Barger:2003zg, Iocco:2008va, Krauss:2010xg}
For a discussion of neutrino asymmetries and oscillations and their impact on BBN c.f.~\cite{Castorina:2012md}.

\section{Used data and method}

To test the theory of standard big bang nucleosynthesis we have two independent possibilities, the analysis of 
the cosmic microwave background (CMB) and the spectral analysis of stellar objects. The baryon to photon 
density $\eta_{10}=10^{10} \eta_b$ given by the final cosmological analysis of the Wilkinson 
Microwave Anisotropy Probe (WMAP 9yr) combined with data from ACT and SPT and priors on the 
baryon acoustic oscillation (BAO) scale and the Hubble expansion rate $H_0$. For a
six-parameter fit to the flat $\Lambda$ cold dark matter model, $\eta_{10} = 6.079 \pm 0.090$ (or 
$\Omega_b h^2 = 0.02223\pm 0.00033$) \cite{Hinshaw:2012fq},
is in agreement with the value from the observation of primordial deuterium of high redshift, 
low metallicity quasi stellar objects, $\eta_{10}(D)=6.0 \pm 0.3$ \cite{Steigman:2007xt}. 

The stellar abundance of deuterium is the easiest to trace back to its primordial value. It is the lightest 
bound state and thus burned in all star burning cycles to $^3$He. The observed deuterium abundance 
at any red shift thus provides a robust upper limit on its primordial value. The SBBN prediction from
$\eta_{10}({\rm WMAP 7yr})$ is D/H$=(2.59 \pm 0.15)\times 10^{-5}$ \cite{Coc:2011az}. One seeks to 
observe young, high redshift and low metallicity damped 
Lyman-$\alpha$ systems. Nowadays nine objects can be used, and their inferred mean of 
$\eta_{10}$ depends on the weighting of the results. \cite{Pettini:2008mq} find a value of 
D/H$=(2.59 \pm 0.15)\times 10^{-5}$, where the value reported in \cite{Olive:2012xf} is 
D/H$=(3.05  \pm .22)\times 10^{-5}$, significantly higher than the SBBN value. 
Note that for the latter there is only an overlap with the SBBN prediction within their 2 sigma deviations.
Neglecting the lower abundances observed (as deuterium is destroyed easily) 
would even lead to D/H$=(3.11 \pm 0.21)\times 10^{-5}$, significantly higher then the SBBN prediction 
\cite{Olive:2012xf}.

The evolution of the relic abundances of $^4$He and $^7$Li and the systematic errors in their observations 
are more difficult and introduce possible errors in the determination of primordial abundances.
The SBBN predictions for $^4$He, or equivalently its mass fraction 
$Y_p(\rm SBBN)=0.2476\pm0.0004$ \cite{Coc:2011az}, are in agreement with observations of low 
metallicity HII regions, $Y_p=0.2534\pm0.0083$ \cite{Aver:2011bw}. However, the same data set, but 
using a different analyzing method leads to $Y_p=0.2565\pm0.006$ \cite{Izotov:2010ca}. Both values 
point to higher mean values of $Y_p$ than predicted by SBBN. 

This is also supported by recent CMB experiments. The latest analysis of SPT 2012 and WMAP 
seven year data with data for the baryonic acoustic oscillation (BAO) and measurements of $H_0$, 
with $Y_p$ left as a free parameter gives $Y_p=0.305 \pm 0.024$ \cite{Hou:2012xq}.
The corresponding value from the final WMAP analysis is $Y_p=0.299 \pm 0.027$ 
\cite{Hinshaw:2012fq}.
The tendency of a larger helium fraction in CMB data then predicted by SBBN or 
observed at redshift $z\approx3.5$ is also supported by WMAP data alone and the Atacama Comology 
Telescope \cite{Dunkley:2010ge}.

The direct observations of primordial lithium ($^7$Li/H) differ from the SBBN prediction by a factor 
of 4 to 5. The SBBN prediction is $^7$Li/H$=(5.07^{+0.71}_{-0.62})\times 10^{-10}$ \cite{Cyburt:2008kw} 
and $^7$Li/H$=(5.24\pm0.5)\times 10^{-10}$ \cite{Coc:2011az}. All observational data differ significantly 
from this prediction. \cite{Sbordone:2010zi} found $^7$Li/H$=(1.58\pm0.31)\times 10^{-10}$ for a set of 
halo dwarf stars and \cite{Monaco:2010mm} found $^7$Li/H$=(1.48\pm0.41)\times 10^{-10}$ for 
the abundance in omega centauri. This is the so-called Lithium problem, which might have multiple roots, 
see i.e.~\cite{Cyburt:2008kw,Spite:2012us}. It seems that there are several reasons for a depletion of lithium, 
however a precise understanding of this effect is still missing. We can thus regard the lithium observations 
as robust lower limits on the primordial lithium abundance.  

To constrain further the neutrino chemical potentials, we use the CMB analysis of SPT 2012 
\cite{Hou:2012xq}. For their analysis they used further cosmological data: the effect of baryonic 
acoustic oscillations (BAO), the WMAP 7 year data analysis (WMAP7), and measurements of $H_0$. 
We also use the final WMAP analysis in combination with older ACT and SPT data.

\begin{table}
\centering
{
	\begin{tabular}{l|l|l|l} \hline 
	Parameter       &    Value          &    Dataset        &   Reference \\ \hline \hline 
	$100\Omega_bh^2$ & 2.223$\pm$0.033 	& WMAP9+CMBe+BAO+$H_0$ ($\Lambda$CDM) & 
		\cite{Hinshaw:2012fq} \\
 	 				 & 2.266$\pm$0.049 	& WMAP9+CMBe+BAO+$H_0$ 
					 ($Y_p$, $N_{\rm eff}$ free) & \cite{Hinshaw:2012fq} \\ \hline
 	$Y_p$              & 0.2485 $\pm$ 0.003	&  SBBN ($N_{\nu}=3$)     & \cite{Steigman:2007xt} \\
	                          & 0.2534 $\pm$ 0.0083	& extragalactic HII regions & \cite{Aver:2011bw} \\
	\hline
	(D/H)$\times 10^{5}$ & $2.59 \pm 0.15$	& SBBN for $\Omega_b h^2$(CMB)	& \cite{Coc:2011az} \\
				& $2.82 \pm 0.20$		& quasar absorption lines		& \cite{Pettini:2008mq} \\
				& $3.05 \pm 0.22$		& quasar absorption lines		& \cite{Olive:2012xf} \\ \hline
      ($^7$Li/H)$\times10^{-10}$	 & $5.07^{+0.71}_{-0.62}$ & SBBN for $\Omega_bh^2$(CMB)  &
      		\cite{Cyburt:2008kw}\\
				& $1.48 \pm 0.41$		& Omega Centauri 		& \cite{Monaco:2010mm} \\	
				& $1.58 \pm 0.31$		& halo star abundance	& \cite{Sbordone:2010zi} \\ \hline\hline
	\end{tabular}
}
\caption{Used parameters and values.}
\label{tab:Observations}
\end{table}

To use the constraints from CMB observations on $\xi_f$, we assume for nucleosynthesis calculations 
that the neutrinos are effectively massless at the time of photon decoupling, 
and that they have no interactions with other particles. Thus we can assume 
$\xi_f(T_{\rm CMB})=\xi_f(T_{\rm BBN})$. We also assume that neutrino flavour is in equilibrium 
before the freeze-out of the neutron-to-proton ration at the begin of the BBN epoch.
Assuming all neutrino oscillation parameters fixed to their measured or best-fit values, this is a 
good approximation \cite{Castorina:2012md}.
Our assumptions lead then to one additional parameter for SBBN, a non-vanishing 
$\mu_{\nu_f}(N_{\rm eff})$.

This well motivated extension of the SBBN gives us the possibility to use CMB data alone to measure 
$\eta_{10}$, $Y_p$ and $N_{\rm eff}$ at the same time and thus to constrain $\xi_f$ from the CMB alone. 
In the following, we apply the extension of the standard model with neutrino chemical potentials to predict 
the relative primordial abundances of light elements. We perform our calculations with a full numeric 
BBN code and compare the results to recent reported abundances of light elements, displayed in 
table \ref{tab:Observations}.

We used a modified version of the PArthENoPE code \cite{Pisanti:2007hk} to calculate abundances for varying neutrino chemical
potentials $\xi_f$, equal for all flavours, for a present day baryon to photon density 
$100\Omega_bh^2= 2.266\pm0.049$ from the combined analysis of SPT, ATC and WMAP 9yr including 
BAO and H0, and $Y_p$ and $N_{\rm eff}$ left free and independent of each other \cite{Hinshaw:2012fq}. 

\begin{figure}
\hfill	\includegraphics[angle=270,width=0.8\textwidth]{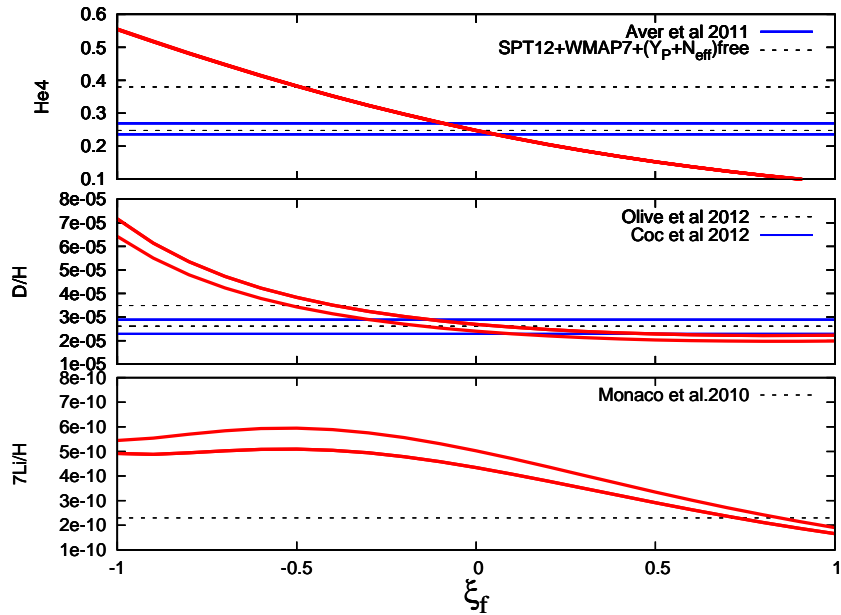}
\caption{Comparison of predicted and measured primordial abundances of helium, deuterium, and lithium. 
The red lines represent the primordial abundances of $^4$He ($Y_p$), D/H and $^7$Li/H as a function 
of neutrino chemical potential $-1< \xi_f < 1$. We use the PArthENoPE code and show the allowed range 
of abundances  corresponding to the 2$\sigma$ region of $100 \Omega_b h^2 = 2.266\pm0.043$  
\cite{Hinshaw:2012fq}. We assume $\xi_e = \xi_\mu = \xi_\tau$. The blue and black dashed lines 
represent various observational constraints on the observed abundance of elements. We do not include 
a guess for lithium depletion.
\label{fig:BBN1}}
\end{figure}

\section{Results}

The results of our calculations are presented in figure \ref{fig:BBN1} and table \ref{tab:results}. 
Comparing observed abundances of  $^4$He, D/H and $^7$Li/H  to BBN abundance predictions, 
we can constrain the neutrino chemical potential. 

\begin{table}
	\centering
	{\small
		\begin{tabular}{l|l|l} \hline 
		Element 		& Allowed Region    		&    Dataset                          				\\ \hline \hline
	 	$^4$He		&$-0.091 < \xi_f < 0.051$ 	& $^4$He (extra galactic HII) \cite{Aver:2011bw}  				\\ 
	 				&$-0.470 < \xi_f < 0.022$	& 
						SPT12+WMAP7  \cite{Hou:2012xq}   	\\
					&$-0.405 < \xi_f < 0.056$	& 
						SPT12+WMAP7+BAO+$H_0$  \cite{Hou:2012xq}  \\
					&$-0.461 < \xi_f < 0.250$	& 
						SPT11+WMAP7  \cite{Keisler:2011aw}  	\\
					&$-0.460 < \xi_f < 0.120$	& 
						WMAP9+ACT11+SPT11 \cite{Hinshaw:2012fq} \\
					&$-0.380 < \xi_f <0.190$	& 
						WMAP9+ACT11+SPT11+BAO+$H_0$  \cite{Hinshaw:2012fq} \\
		D/H 			&$-0.347 < \xi_f <0.153$	& quasar absorption lines	\cite{Pettini:2008mq}	   	\\
					&$-0.524 < \xi_f <0.055$	& quasar absorption lines	 \cite{Olive:2012xf}  	\\
		$^7$Li/H 		&$0.700 < \xi_f$		& Omega Centauri (without depletion)
						 \cite{Monaco:2010mm}	\\
					&$0.767 <\xi_f$		& halo star abundance (without depletion) 
						\cite{Sbordone:2010zi}	\\\hline\hline
		\end{tabular}
	}
	\caption{Constraints on $\xi_f$ from different observational data. All reported CMB results rely on fits 
	to a 8-parameter flat $\Lambda$CDM model, with $Y_p$ and $N_{\rm eff}$. Note that our constraints on $^7$Li/H are limited to our calculations $|\xi_f|<1$.}
	\label{tab:results}
\end{table}

\subsection{Helium-4}

For the $^4$He abundance observed in extragalactic HII regions \cite{Aver:2011bw} we find an 
allowed 2$\sigma$ region of $-0.091 < \xi_f < 0.051$. The helium abundance inferred from the CMB is 
somewhat higher. The combined analysis of the recent SPT and WMAP 7 year data \cite{Hou:2012xq} 
in a $\Lambda$CDM model with $Y_p$ and $N_{\rm eff}$ as free parameter and independent of each 
other allows for $-0.470 < \xi_f <0.022$. Note, that allowed positive chemical potentials are smaller than the 
ones referred from the HII regions. Including further BAO and $H_0$ data leads to $-0.405<\xi_f<0.056$. 
The allowed range of negative chemical potentials  is 4 times larger compared to the constraints from 
the extragalactic HII regions, while it is more or less the same for the positive chemical potentials.  
This trend is also supported by the WMAP 9 year analysis. This final analysis is performed with older SPT 
data from \cite{Keisler:2011aw} and ACT data from \cite{Dunkley:2010ge} for that we also provide the 
ranges from that analysis. Thus the analysis of all recent CMB data sets allows 
for negative chemical potentials four to five times larger than observations of primordial helium in 
extragalactic HII regions suggest. Interestingly, the strongest limit on positive chemical potentials stems 
from the combination of WMAP7 and SPT12 data.   
These bounds also demonstrate the potential of upcoming CMB data releases to further constrain $\xi_f$.    

\subsection{Deuterium} 

For the deuterium abundance we found an interesting difference for the 2$\sigma$ overlap of the two 
concurring observational mean values with our calculation
(see figure 1). For the measurement of D/H of \cite{Pettini:2008mq}, $-0.35<\xi_f<0.15$ is allowed. 
From the data analysis of \cite{Olive:2012xf}, we find $-0.52<\xi_f<0.05$. 
The upper bound is almost identical with the one from $^4$He abundance of extragalactic HII regions. If we 
compare our results to the higher 
D/H$=(3.11 \pm 0.21)\times 10^{-5}$ reported in \cite{Olive:2012xf}, we find $-0.51<\xi_f<-0.02$. 
In this case the deuterium abundance would exclude positive neutrino chemical potentials at the 95\%~C.L.

\subsection{Lithium} 

In the bottom panel of figure 1 we confront our calculation for the $^7$Li/H with observations. We found an 
agreement of theory and observed abundance for large positive neutrino chemical potentials 
$0.7<\xi_f$ for \cite{Monaco:2010mm} and $0.77<\xi_f$ for the abundance of \cite{Sbordone:2010zi}.
However, neutrino chemical potentials that large are excluded by helium-4, deuterium and CMB 
data.  On the other hand, it is very plausible that lithium has been depleted in the course of the galactic 
chemical evolution \cite{Burke:2003bj}.  Thus we conclude, that even a large neutrino asymmetry of the
Universe, would not solve the lithium problem. As in the case of SBBN, a lithium depletion by a 
factor of at least 2 to 3 is required.

\subsection{A consistent picture?} 

Bringing everything together, the most stringent individual constraint on the neutrino 
chemical potential stems from the helium-4 abundance from extragalactic HII regions, 
$-0.09<\xi_f^{\rm He} <0.05$. The combined analysis of SPT12 and WMAP7 data provides 
an even more stringent upper bound on positive neutrino chemical potentials from the helium 
abundance, $\xi_f^{\rm CMB} < 0.02$. The resulting range $-0.09<\xi_f^{\rm He, CMB} <0.02$ is consistent 
with all CMB data and with the observed abundance of deuterium. The lithium problem remains.          

However, one should keep in mind, that the observations of HII regions may not be as representative 
for the Universe as the helium-4 abundance inferred from the CMB.
If we rely on the helium-4 abundance from the CMB and combining it with the deuterium abundance 
(taking also systematic uncertainties into account), leads to a range
$-0.38 < \xi_f^{\rm CMB, D} < 0.02$, consistent with all data sets.
Even for this increased interval, the lithium problem remains. 
In both situations, negative values of $\xi_f$ are preferred.

\begin{figure}
\hfill \includegraphics[angle=0,width=0.42\textwidth]{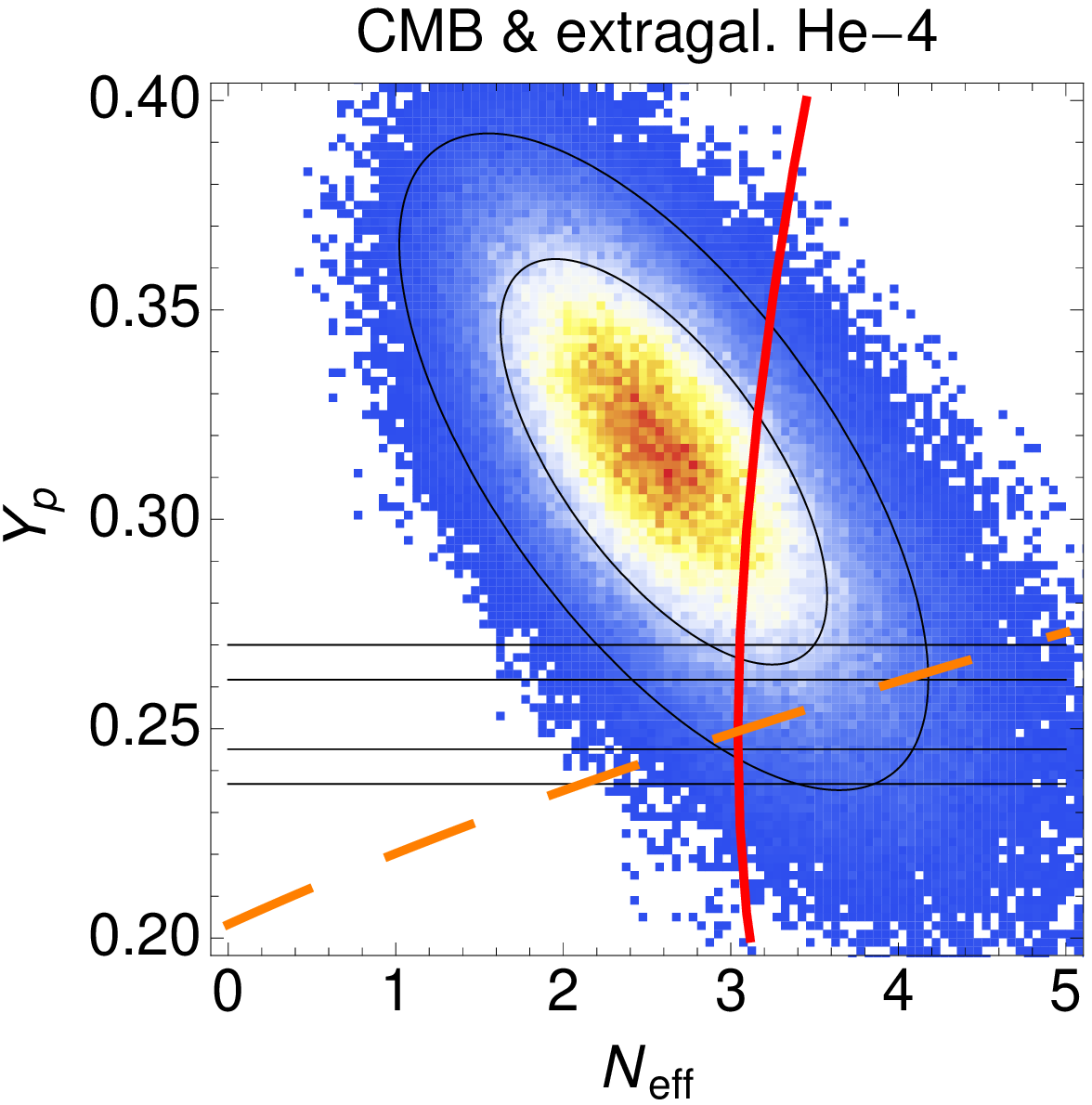}
\includegraphics[angle=0,width=0.42\textwidth]{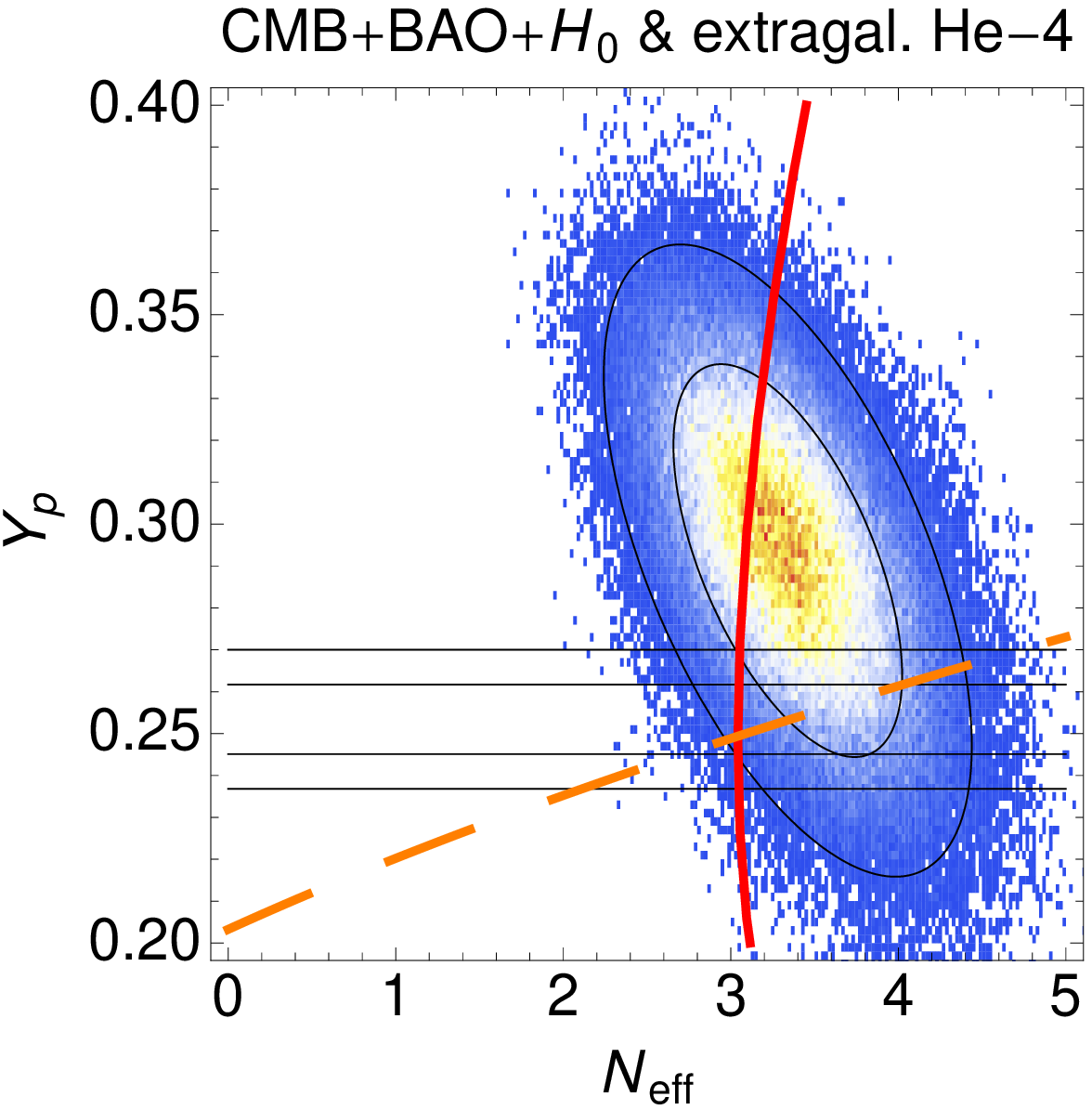}
\caption{The number of relativistic degrees of freedom versus primordial abundance of helium-4. The full red 
line describes BBN with a neutrino chemical potential. The dashed orange line describes BBN with extra 
relativistic degrees of freedom. The crossing point corresponds to SBBN. 
Negative chemical potentials, $\xi_f < 0$, 
increase the abundance of helium-4. The ellipses denote the 68\% and 95\% contours of the 
Markov chains of the SPT and WMAP 7 year analysis in \cite{Hou:2012xq}, superposed 
on a density histogram of the chains. The horizontal thin lines denote the 68\% and 95\% contours of 
the helium-4 measurement of extragalactic HII regions 
\cite{Aver:2011bw}.  The left panel uses CMB data and direct abundance measurements only, in the right panel data from BAO and $H_0$ measurements have been added. \label{fig2}}
\end{figure} 
 
\section{Conclusion}

Recent CMB data, combined with priors obtained from BAO data and measurements of $H_0$, 
point to a high helium fraction compared to standard BBN and observed in HII regions.
At the same time there might be some extra radiation degrees of freedom, expressed in $N_{\rm eff}$. 

Introducing a single additional variable to the standard model 
of cosmology, a non-vanishing neutrino chemical potential induced by a large lepton asymmetry, leads 
naturally to higher primordial helium without affecting the abundance of primordial light elements too much.

Allowing for the helium fraction $Y_p$ and $N_{\rm eff}$ to be free parameters in the analysis of 
CMB data, gives, for the combination WMAP9+ACT11+SPT11+BAO+$H_0$, 
$Y_p = 0.278^{+0.034}_{ -0.032}$ and $N_{\rm eff} = 3.55^{+0.49}_{-0.48}$ \cite{Hinshaw:2012fq}. 
Also CMB alone (SPT12+WMAP7) points to a higher $Y_p = 0.314 \pm 0.033$ and $N_{\rm eff} = 
2.60 \pm 0.67$ \cite{Hou:2012xq}.
In \cite{Hou:2012xq} and  \cite{Hinshaw:2012fq} these findings have been 
compared to SBBN with and without extra relativistic degrees of freedom. We show a similar plot in 
figure \ref{fig2}. Here we compare the Markov chains from \cite{Hou:2012xq} with the BBN prediction 
for two different scenarios. The thick red line is the scenario considered in this work, i.e.~BBN 
with $\xi_f \neq 0$ and $N_\nu = 3.042$. The thick dashed orange line is the scenario with extra 
radiation degrees of freedom and $\xi_f = 0$, e.g.~the case of one or two sterile neutrinos. 
The crossing point of the two models is the SBBN prediction. The part of the thick red line above this 
crossing corresponds to $\xi_f < 0$. Thus a negative chemical potential would allow for a BBN best-fit 
model much closer to the maximum of the CMB posterior distribution of $Y_p$ and 
$N_{\rm eff}$. The chemical potential seems to do better than just adding extra degrees of freedom.

As was shown in \cite{Hou:2012xq} the analysis of the CMB data combined with BAO and H0 shows 
a preference for an extension of the $\Lambda$CDM model with $N_{\rm eff}$ and massive neutrinos. 
The CMB alone favours the one parameter extension of including a running of the spectral index of 
primordial density perturbations or a two-parameter extension with $Y_p$ and $N_{\rm eff}$ as free 
parameters. The introduction of a neutrino chemical potential has the advantage that it can improve the 
fit to both data sets with only a single additional parameter.
 
Here we suggested that recent CMB data could provide a first hint towards
a lepton asymmetry of the Universe, much larger than the baryon asymmetry of the Universe. 
Today this lepton asymmetry would hide in the neutrino background. This scenario would have 
interesting implications for the early Universe, especially at the epochs of the cosmic quark-hadron transition 
\cite{Schwarz:2009ii} and WIMP decoupling \cite{Stuke:2011wz}. 
The largest allowed ($2\sigma$) neutrino chemical potential $|\xi_f|=0.45$ leads to 
$\Delta N_{\rm eff} = 0.266$, fully consistent with CMB observations. 

The helium fraction reported by CMB observations results in a negative neutrino chemical potential 
$\xi_f \sim - 0.2$ and $\Delta N_{\rm eff} \sim 0.1$. In that case we would live in a Universe ruled by 
anti-neutrinos. From our analysis we conclude that the present abundance of light elements and 
CMB data are not able to rule out $\xi_f = 0$, the standard scenario of BBN. However, upcoming 
CMB data releases and improved measurements of primordial abundances will allow us to test the 
idea of a leptonic Universe.

\section*{Acknowledgement}

We thank Glenn Starkman for valuable comments and discussions during an early stage of this project. We acknowledge the use of the PArthENoPE code. 
Parts of this work have been financially supported by the Deutsche Forschungs Gemeinschaft under grant GRK881.

%%%%%%%%%%%%%%%%%%%%%%%%%%%%%%%%%%%%%%%%%%%%%%%%%%%%%%%%%%%%%%%%%%%%%%%%%%%%%%%%%%%%%%%%%%%%%%%%%
%bibliography

\end{document}